\documentclass[english]{IEEEtran}
\usepackage[T1]{fontenc}
\usepackage[latin9]{inputenc}
\usepackage[a4paper]{geometry}
\usepackage{amsmath}
\usepackage{amssymb}
\usepackage{stackrel}
\usepackage{graphicx}

\makeatletter
\usepackage{colortbl}
\usepackage{cite}
\renewcommand{\figurename}{Fig.}
\usepackage{nomencl}
\usepackage{stfloats}
\usepackage[margin=8pt,font=footnotesize,labelfont=md,labelsep=colon]{caption}

\usepackage{multicol}
\usepackage{float}
\usepackage{lipsum}
\makeatother

\usepackage{babel}
\begin{document}

\title{On the Performance of DF-based Power-Line/Visible-Light Communication
Systems }

\author{Waled Gheth, Khaled M. Rabie, Bamidele Adebisi, Muhammad Ijaz and
Georgina Harris \\
School of Engineering, Manchester Metropolitan University, Manchester,
UK\\
Emails:\{w.gheth, k.rabie, b.adebisi, m.ijaz, g.harris\}@mmu.ac.uk }

\maketitle
\thispagestyle{empty}
\pagestyle{empty}
\begin{abstract}
This paper presents a comprehensive performance analysis of an integrated
indoor power line communication (PLC)/visible light communication
(VLC) system with the presence of a decode-and-forward (DF) relay.
The existing indoor power line networks are used as the backbone for
VLCs. The performance of the proposed system is evaluated in terms
of the average capacity and the outage probability. A new unified
mathematical method is developed for the PLC/VLC system and analytical
expressions for the aforementioned performance metrics are derived.
Monte Carlo simulations are provided throughout the paper to verify
the correctness of the analysis. The results reveal that the performance
of the proposed system deteriorates with increasing the end-to-end
distance and improves with increasing the relay transmit power. It
is also shown that the outage probability of the system under consideration
is negatively affected by the vertical distance to user plane. 
\end{abstract}


\section{Introduction and Related Works}

\IEEEPARstart{R}{eliable}  communications can be achieved by integrating
different networks such as power line communication (PLC) with visible
light communication (VLC) and PLC with wireless communication systems.
Such systems have been studied by many researchers and have shown
very promising solutions in terms of capacity, security and data rate
improvements \cite{Leo2017,T.Komine2003}. Exploiting the existing
power line networks in buildings and utility grids makes such networks
one of the competitive technologies for broad-band communications
in indoor applications. The other technology, which is expected to
play an essential role in such applications, is the optical wireless
communication (OWC). VLC is one of the indoor OWC technologies which
has attracted a considerable attention because of the widespread usage
of the LEDs. 

Recently, the adoption of other communication technologies for PLC
and VLC systems has motivated many researchers to implement different
relaying protocols to further enhance the system performance, the
most common of which are decode-and-forward (DF), amplify-and-forward
(AF), incremental DF and selective DF relaying, see e.g., \cite{Tan2011,D.lessandro2012,Gupta17,K.M.Rabie,K.R,Rabie18}.

The implementation of half duplex time division relaying protocols
with indoor PLC applications has been investigated by the authors
in \cite{D.lessandro2012}. The authors in \cite{Zou2009} concluded
that the relay deployment in wireless systems is by far more efficient
than that in PLC systems. Two-way relaying and one-way relaying were
discussed in \cite{Tan2011} where the authors reported that there
is a significant improvement in system performance by utilizing the
former compared to the latter. The authors in \cite{Moslem2016} investigated
the deployment of multi-way relaying in parallel for indoor PLC applications.
It was found that the achievable data rates were relatively high compared
to those accomplished by single transmission. The study in \cite{Leo2017}
discussed the hybrid PLC/wireless system performance in the presence
of AF relaying where the source of the information simultaneously
transmits data via both networks PLC and wireless. It was reported
that the integrated PLC/wireless system outperforms the PLC-only and
the wireless-only systems. The performance of cooperative free space
optics (FSO)-VLC communication system in terms of capacity with a
DF relay was investigated in \cite{Gupta17}. The FSO link was the
backbone of the VLC system and the DF relay was located in-between
the two links. It was concluded that this system has high efficiency
in terms of data rate. The cooperation between light sources and how
it can improve the performance of VLC systems was investigated in
\cite{Refik15} where the authors reported that the implementation
of DF relaying offers better performance than that of AF relaying. 

To the best of the authors knowledge, only few works have studied
the performance of cascaded PLC/VLC links in the presence of DF relaying
protocols in terms of capacity and outage probability. Therefore,
this paper investigates the implementation of DF relaying to connect
PLC and VLC links via a DF relay. In the proposed system, the information
source first sends the information signal to a DF relay through a
PLC link. Secondly, the DF relay decodes and forwards the received
signal to the destination node via the VLC channel. The major contribution
of this work resides in deriving analytical expressions for the average
capacity and the outage probability of the proposed hybrid PLC-VLC
DF-based system. Formulating the overall capacity and outage probability
of the proposed PLC/VLC system offers the opportunity to examine the
effect of various system parameters on the communication performance. 

The rest of the paper is organized as follows. The system model is
described in Section \ref{sec:System-Model-1}. Section \ref{sec:Performance-Analysis-1}
presents the system performance analysis. Discussions of the numerical
results are presented in Section \ref{sec:Numerical-Results-1}. Finally,
the conclusions of this paper are drawn in Section \ref{sec:Conclusion-1}.

\section{\label{sec:System-Model-1}System Model}

The system model under consideration is shown in Fig. 1 where PLC
users can be connected to the network through PLC modems. We assume
that the source node and the VLC user are in a different floors. The
DF relay receives the data from the source node via the PLC link.
The data is then decoded and re-transmitted by the DF relay to the
end user through the VLC link. The complex channel gains $h_{P}$
and $h_{v}$ represent the source-to-relay, (i.e., the PLC link) and
the relay-to-destination, (i.e., the VLC link) channel gains, respectively.
Both channels are assumed to be independent and identically distributed.
The channel distribution of the PLC link is log-normal distribution
\cite{Tonello2012}. In this study, we consider the line-of-sight
(LOS) component of the down-link transmission of the LED only as it
represents more than $90\%$ of the total received signal \cite{Zego2009}.
The LED is located on the ceiling with a vertical distance to the
user plane $L$ and Euclidean distance to the destination $d_{k}$.
The VLC channel is subjected to a random distribution which is affected
by the uniform distribution of the user`s location \cite{Zego2009}.
Because of the nature of the network structure adopted here, it is
logical to assume that there is no direct link between the source
and destination nodes. It is worth mentioning, for simplicity and
without loss of generality, that noise over the two links is assumed
to be additive white Gaussian noise (AWGN). 

\section{\label{sec:Performance-Analysis-1}Performance Analysis}

\subsection{Average Capacity}

The instantaneous capacity of the hybrid system is given by the following 

\begin{equation}
C=\textrm{min}(C_{PLC},C_{VLC}),\label{eq:C.total}
\end{equation}
where $C_{PLC}$ and $C_{VLC}$ are the instantaneous capacities of
the PLC and VLC links, respectively.

Therefore, each link capacity should be derived in order to find the
overall capacity. We start with the PLC link which connects the transmitter
with the DF relay and acts as a backhaul to the VLC link. The signal
at the relay can be given by 

\begin{equation}
y_{r}\left(t\right)=\sqrt{P_{s}}e^{-\alpha d}h_{PLC}s(t)+n_{r},\label{eq:Yplc-1-1}
\end{equation}

\noindent where $P_{s}$ represents the source transmit power, $d_{1}$
represents the source-to-relay distance, $s(t)$ is the information
signal with E{[}s{]}=1, $n_{r}$ is the noise at the relay which is
assumed to be complex Gaussian with zero mean and variance $\sigma_{r}^{2}$,
and $\alpha$ is the PLC channel attenuation factor given by $\alpha=a_{0}+a_{1}f^{k}$,
where $a_{0}$ and $a_{1}$ are constants determined by measurements,
$f$ is the system operating frequency and $k$ denotes the exponent
of the attenuation factor \cite{GhethCSNDSPNew}.

\renewcommand{\figurename}{Fig.}  

\begin{figure}
\begin{centering}
\includegraphics[scale=0.1]{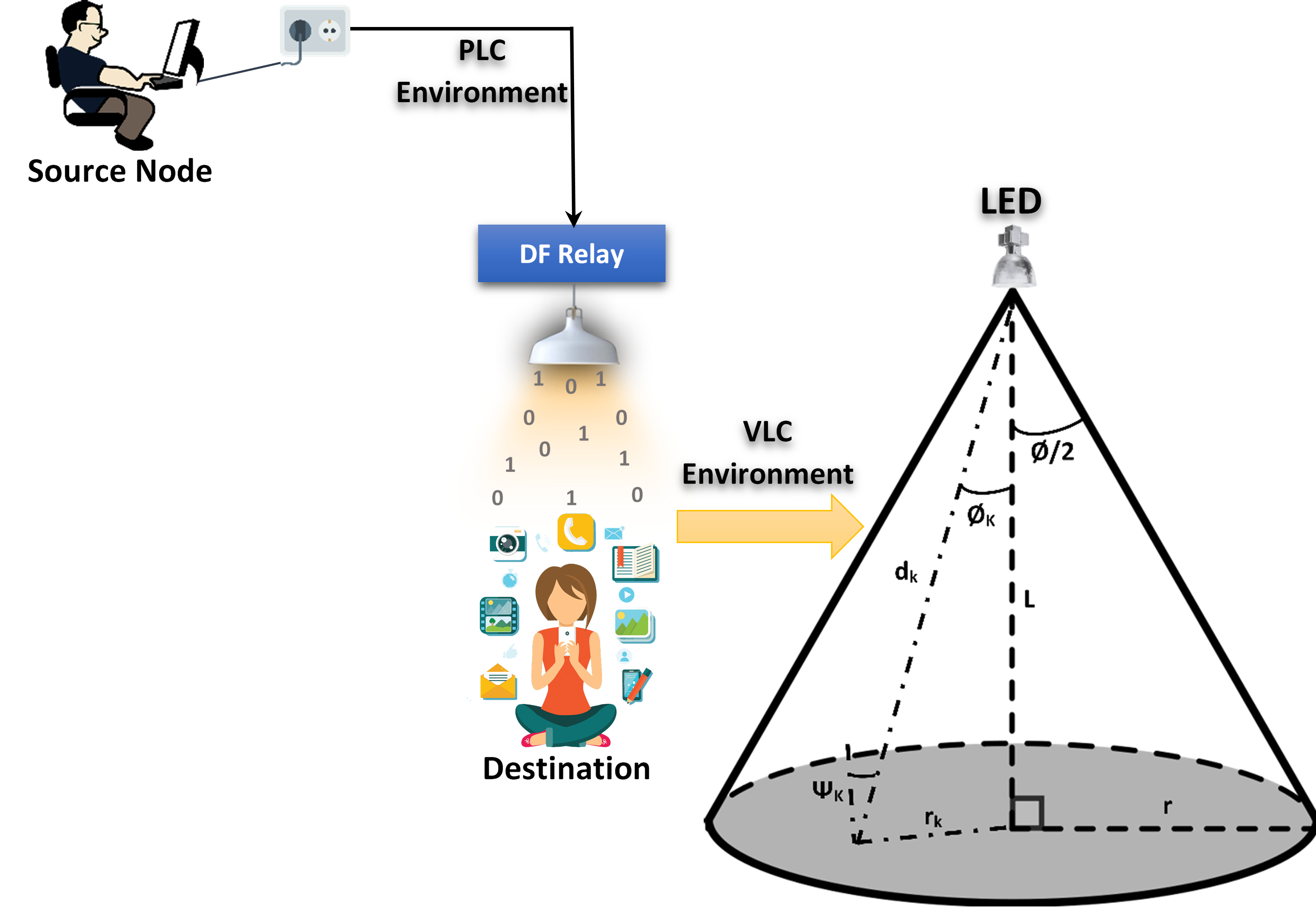}
\par\end{centering}
\caption{System model for the hybrid PLC/VLC network for indoor communication
applications.}

\end{figure}

The signal-to-noise ratio (SNR) at the DF relay can be written as
follows 

\begin{equation}
\gamma_{r}=\frac{P_{s}e^{-2\alpha d}\left|h_{PLC}\right|^{2}}{\sigma_{r}^{2}}.\label{eq:SNRplc-1-1}
\end{equation}

Mathematically, we can calculate the average capacity of the PLC link
as \cite{GhethCSNDSPNew} 

\begin{equation}
\mathbb{E}[C_{PLC}]={\displaystyle {\textstyle \stackrel[n=1]{N_{p}}{\sum}\frac{1}{\sqrt{\pi}}H_{x_{n}}h(x_{n})}},\label{eq:Cplast-1-1}
\end{equation}

\noindent where $H_{x_{n}}$ represents the weight factors of the
$N_{p}$ order Hermite polynomial, $x_{n}$ is the zeros of the same
order and

\begin{equation}
h(x_{n})=\textrm{\textrm{lo\ensuremath{g_{2}}}}\left(1+\exp\left(\frac{\sqrt{8}\sigma x_{n}+2\mu+\zeta ln(a)}{\zeta}\right)\right),\label{eq:lastone-1-1}
\end{equation}

\noindent where $\zeta$ is the scaling constant and is equal to $10/\textrm{ln}(10)$,
$\mu$ and $\sigma$ are the log-normal parameters, and $a=\frac{P_{s}e^{-2\alpha d}}{\sigma_{d2}^{2}}.$ 

\noindent Now we calculate the average capacity of the VLC link. The
received signal at the VLC user can be expressed as 

\begin{equation}
y_{d}\left(t\right)=s_{2}(t)h_{v}\sqrt{P_{r}}+n_{d},\label{eq:impulsivenoise2-1}
\end{equation}

\noindent where $s_{2}(t)$ is the information signal with E{[}s{]}=1,
$P_{r}$ is the relay transmit power and $n_{d}$ represents the destination
noise with zero mean and variance $\sigma_{d}^{2}$. 

The SNR at the destination can be written as 

\begin{equation}
\gamma_{d}=\frac{G^{2}P_{r}\left|h_{v}\right|^{2}}{\sigma_{d}^{2}}.\label{SNRvlc}
\end{equation}

The average capacity of the VLC link is calculated as 

\begin{equation}
\mathbb{E}[C_{VLC}]=\stackrel[t_{min}]{{\displaystyle t_{max}}}{\int}\textrm{\ensuremath{\mathrm{log_{2}}}}\left(1+\gamma\right)f(\gamma)d\gamma,\label{Cvlc}
\end{equation}
where $f(\gamma)$ is the probability density function (PDF) , $t_{min}=\frac{\left(Q\left(m_{k}+1\right)L^{m_{k}+1}\right)^{2}}{(r^{2}+L^{2})^{m_{k}+3}}$
and $t_{max}=\frac{\left(Q\left(m_{k}+1\right)L^{m_{k}+1}\right)^{2}}{L^{2(m_{k}+3)}}$.

Considering the random distribution mentioned above for the VLC link
and the Lambertian radiation pattern for LED light emission, the VLC
channel gain $h_{v}$ can be written as 

\begin{equation}
h_{v}=\frac{m_{k}+1}{2\pi d_{k}^{2}}A_{d}cos^{m_{k}}(\phi)cos\left(\Psi_{K}\right)U\left(\Psi_{K}\right)g\left(\Psi_{K}\right)R_{p},\label{eq:Hv}
\end{equation}

\noindent where $A_{d}$ is the detection area of the detector, $d_{k}=\sqrt{r_{k}^{2}+L^{2}}$,
$U\left(\Psi_{K}\right)$ represents the the optical filter gain,
$g\left(\Psi_{K}\right)$ denotes the optical concentration gain,
the responsivity of the photo-detector is represented by $R_{p}$,
$\phi$ is the total angel of the LED, $cos^{m_{k}}(\phi)=cos\left(\Psi_{K}\right)=\frac{L}{\sqrt{r_{k}^{2}+L^{2}}}$,
and $m_{k}$ is the order of the Lambertian radiation pattern which
is given by 

\begin{equation}
m_{k}=\frac{-1}{\textrm{lo\ensuremath{g_{2}}}(cos(\phi/2))},\label{eq:mk1}
\end{equation}

\noindent where $\phi/2$ denotes the semi-angle of the LED.

To simplify our analysis, let us assume that $Q=\frac{1}{2\pi}A_{d}U\left(\Psi_{K}\right)g\left(\Psi_{K}\right)R_{p}$.
Hence, (\ref{eq:Hv}) can be rewritten as 

\begin{equation}
h_{v}=\frac{Q(m_{k}+1)L^{m+1}}{(r_{k}^{2}+L^{2})^{^{\frac{m+3}{2}}}}.\label{eq:Hv2}
\end{equation}

It is assumed that the location of the users is uniformly distributed
with PDF given as 

\begin{equation}
f_{r_{k}}(r)=\frac{2r}{r^{2}}.
\end{equation}

The PDF of the un-ordered channel gain of the VLC link can be driven
using the change-of-variable method used in \cite{Gupta17} as follows 

\begin{equation}
f_{h_{k}}(h)=\mid\mathfrak{\frac{d}{d\mathit{h}}\mathit{u^{-1}(h)}}\mid f_{h_{k}}(u^{-1}(h)).
\end{equation}

We can now derive the PDF of the VLC channel gain given by

\begin{equation}
f_{h_{k}}=\frac{2Q^{\frac{2}{2+m}}\left(\left(m_{k}+1\right)L^{m_{k}+1}\right)^{\frac{2}{m+3}}h^{-\frac{2}{m_{k}+3}-1}}{(m_{k}+3)r^{2}},\label{eq:Fhk}
\end{equation}

\noindent where $r$ is the maximum cell radius of the LOS.

The PDF of $h_{k}^{2}$ can now be obtained as 

\begin{equation}
f_{h_{k}^{2}}=\frac{-Q^{\frac{2}{2+m}}\left(\left(m_{k}+1\right)L^{m_{k}+1}\right)^{\frac{2}{m+3}}h^{-\frac{m_{k}+4}{m_{k}+3}}h^{\frac{1}{m_{k}+3}}}{(m_{k}+3)r^{2}}.\label{eq:Fvlc}
\end{equation}

Substituting (\ref{SNRvlc}) and (\ref{eq:Fvlc}) in (\ref{Cvlc}),
we obtain (\ref{eq:Cvlc}), shown at the top of the next page,

\begin{figure*}
\begin{equation}
\mathbb{E}[C_{VLC}]=\frac{1}{2\textrm{log\ensuremath{\left[2\right]}}}\stackrel[t_{min}]{{\displaystyle t_{max}}}{\int}\textrm{log\ensuremath{\left[1+\mathit{\frac{P_{r}}{\sigma_{d}^{2}}h}\right]}}\left(\frac{-Q^{\frac{2}{2+m}}\left(\left(m_{k}+1\right)L^{m_{k}+1}\right)^{\frac{2}{m+3}}h^{-\frac{m_{k}+4}{m_{k}+3}}h^{\frac{1}{m_{k}+3}}}{(m_{k}+3)r^{2}}\right).\label{eq:Cvlc}
\end{equation}
\centering{}\rule[0.5ex]{2\columnwidth}{0.8pt}
\end{figure*}

Substituting $t_{min}$, $t_{max}$ values in (\ref{eq:Cvlc}) we
obtain (\ref{eq:finalone}), shown at the top of the next page.

\begin{figure*}
\begin{align*}
\mathbb{E}[C_{VLC}] & =\frac{\frac{1}{(m_{k}+3)r^{2}}\left(Q\left(m_{k}+1\right)L^{m_{k}+1}\right)^{\frac{2}{m+3}}}{2\textrm{log\ensuremath{\left[2\right]}}}(m_{k}+3)\\
 & \left(t_{max}^{-\frac{1}{m_{k}+3}}\left(-3-m_{k}+\left(3+m_{k}\right)_{2}F_{1}\left[1,-\frac{1}{m_{k}+3},\frac{m_{k}+2}{m_{k}+3},-t_{max}\frac{P_{r}}{\sigma_{d}^{2}}\right]-\textrm{log}\left[1+t_{max}\frac{P_{r}}{\sigma_{d}^{2}}\right]\right)\right.
\end{align*}
\begin{equation}
\begin{split} & \left.\qquad\qquad-t_{min}^{-\frac{1}{m_{k}+3}}\left(-3-m_{k}+\left(3+m_{k}\right)_{2}F_{1}\left[1,-\frac{1}{m_{k}+3},\frac{m_{k}+2}{m_{k}+3},-t_{min}\frac{P_{r}}{\sigma_{d}^{2}}\right]-\textrm{log}\left[1+t_{min}\frac{P_{r}}{\sigma_{d}^{2}}\right]\right)\right).\end{split}
\label{eq:finalone}
\end{equation}
\centering{}\rule[0.5ex]{2\columnwidth}{0.8pt}
\end{figure*}

\subsection{Outage Probability }

The outage probability is simply defined as the probability that the
instantaneous SNR of system is less than a certain threshold value,
$R_{th}$, and is given as 

\begin{equation}
P_{outage}=P_{r}\left[C<R_{th}\right].
\end{equation}

The end-to-end outage probability for the proposed system can be written
as 

\begin{equation}
P_{outage}=P_{out}^{PLC}+\left(1-P_{out}^{PLC}\right)P_{out}^{VLC},
\end{equation}
where $P_{out}^{PLC}$is outage probability of the first link and
$P_{out}^{VLC}$ is the outage probability of the second link.

The cumulative density function (CDF) of each link represents the
outage probability of that link. Therefore, the outage probability
of the PLC and VLC links can be calculated as 

\begin{equation}
P_{out}^{PLC}=\stackrel[0]{{\displaystyle \infty}}{\int}\frac{\zeta}{z\sqrt{8\pi}\sigma}\textrm{exp}\left(-\frac{(\zeta\textrm{ln}(\gamma)-(2\mu+\zeta\textrm{ln}(a_{1})))^{2}}{8\sigma^{2}}\right)d_{z},
\end{equation}

\noindent and

\begin{equation}
P_{out}^{VLC}=\frac{-1}{r^{2}}\left(\left(m_{k}+1\right)QL^{m_{k}+1}\right)^{\frac{2}{m_{k}+3}}h^{\frac{-1}{m_{k}+3}}+\left(1+\frac{L^{2}}{r^{2}}\right).
\end{equation}

\noindent 
\begin{figure}
\centering{}%
\begin{minipage}[t]{0.52\columnwidth}%
\begin{center}
\includegraphics[width=1\columnwidth]{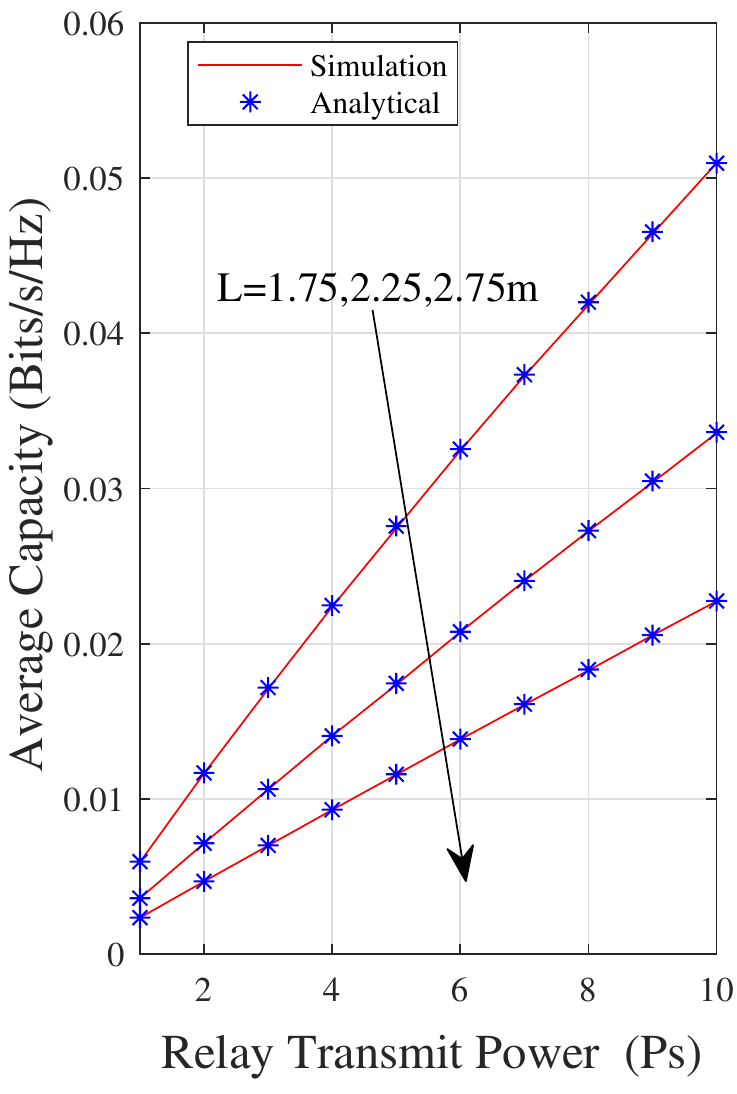}
\par\end{center}
\caption{Average capacity as a function of relay transmit power for different
values of the vertical distance to the user plane.\label{fig:Pr and L} }
\end{minipage}%
\begin{minipage}[t]{0.52\columnwidth}%
\begin{center}
\includegraphics[width=1\columnwidth]{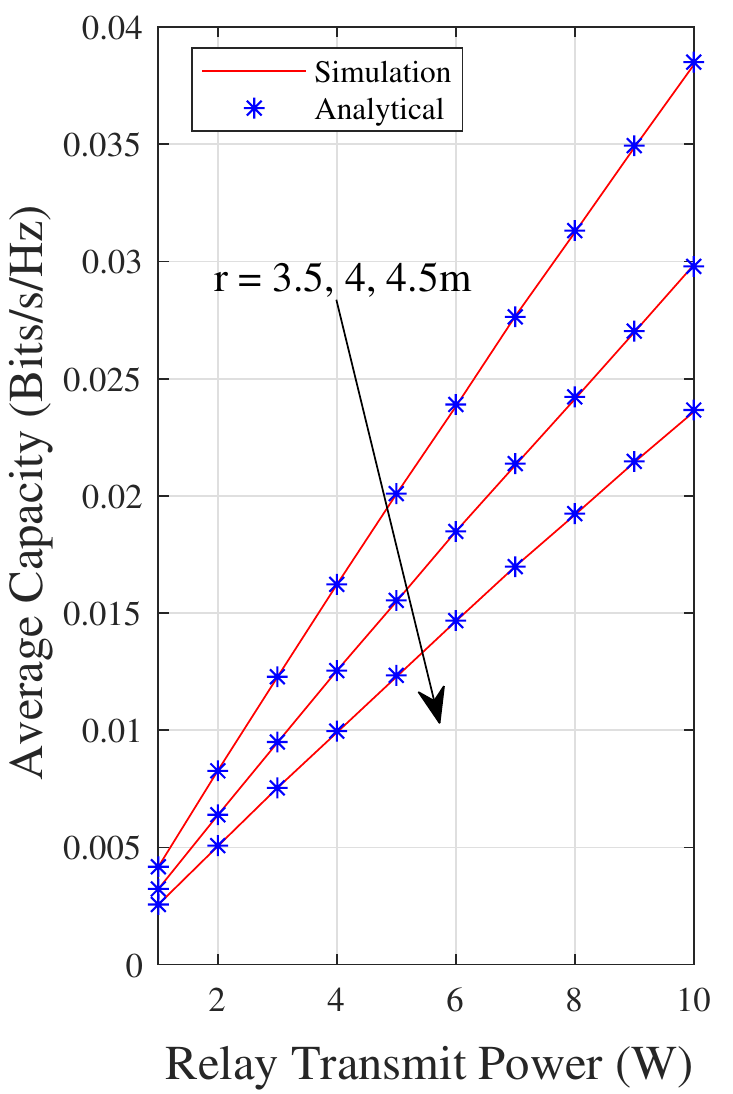}
\par\end{center}
\begin{center}
\caption{Average capacity with respect to the relay transmit power for different
values of the maximum cell radius of the LOS.\label{fig:Pr and Re}}
\par\end{center}%
\end{minipage}
\end{figure}

\section{\label{sec:Numerical-Results-1}Numerical Results}

This section presents some numerical results of the derived expressions
above. The system parameters under consideration are, unless indicated
otherwise, as follows: operating frequency of the system $f=500$kHz,
$k=0.7$, $a_{0}=2.03\times10^{-3}$, $a_{1}=3.75\times10^{-7}$,
$d_{1}=30\textrm{m}$, the input power $P_{s}=0.1$W, $P_{r}=0.1$W,
input SNR is 10dB, $A_{d}=0.1$m, $U\left(\Psi_{K}\right)=g\left(\Psi_{K}\right)=7$dB,
$R_{p}=0.4$A/W, $r=3.6$m, $L=2.15$m and $\phi/2=60^{\circ}$ \cite{GHETH_GLOBECOM}. 

The effect of the relay transmit power and the vertical distance to
the user plane on the performance of the proposed hybrid system is
presented in Fig. \ref{fig:Pr and L}. It is clear that the analytical
results, obtained from (\ref{eq:finalone}), are in perfect agreement
with the simulation ones. As it can be seen, the performance improves
as the relay transmit power increases and/or the vertical distance
of the LED decreases. Fig. \ref{fig:Pr and Re} illustrates the impact
of the maximum cell radius of the LOS on the system performance. It
is noticeable that the performance of the system has a noticeable
improvement when the maximum cell radius of the LOS decreases, particularly
for the high values of the relay transmit power. 

\begin{figure}
\centering{}%
\begin{minipage}[t]{0.52\columnwidth}%
\begin{center}
\includegraphics[width=1\columnwidth]{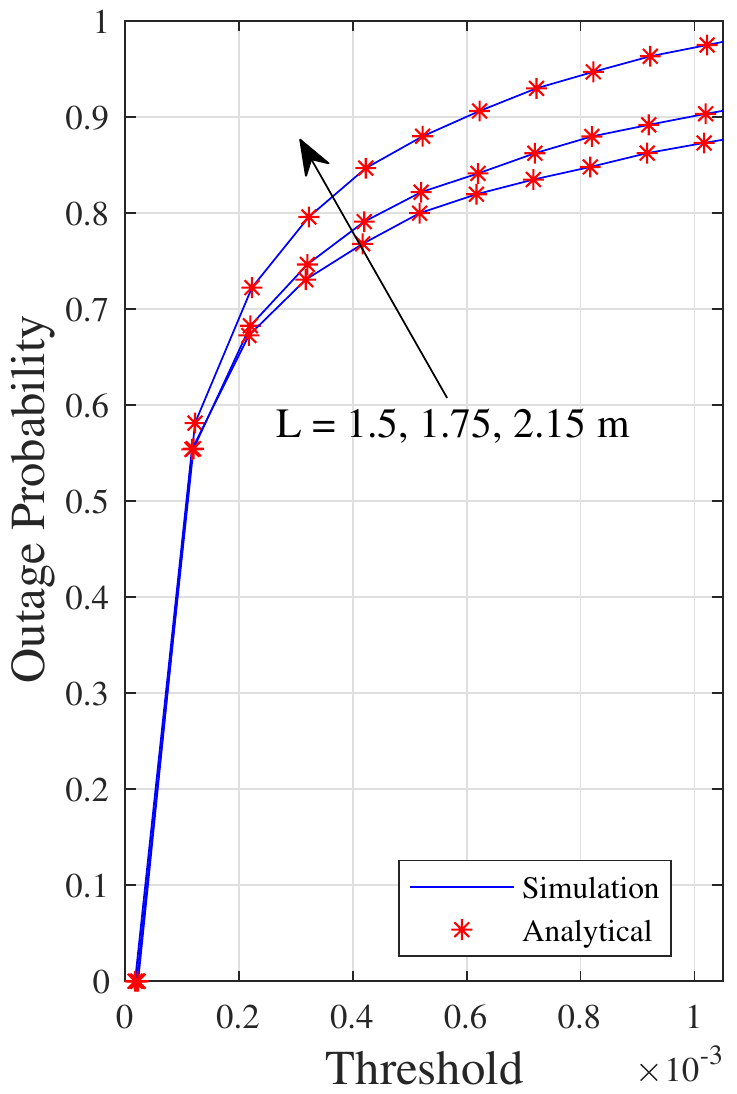}
\par\end{center}
\caption{Outage probability versus the threshold value for different values
of the vertical distance to the user plane.\label{fig:OutageL} }
\end{minipage}%
\begin{minipage}[t]{0.52\columnwidth}%
\begin{center}
\includegraphics[width=1\columnwidth]{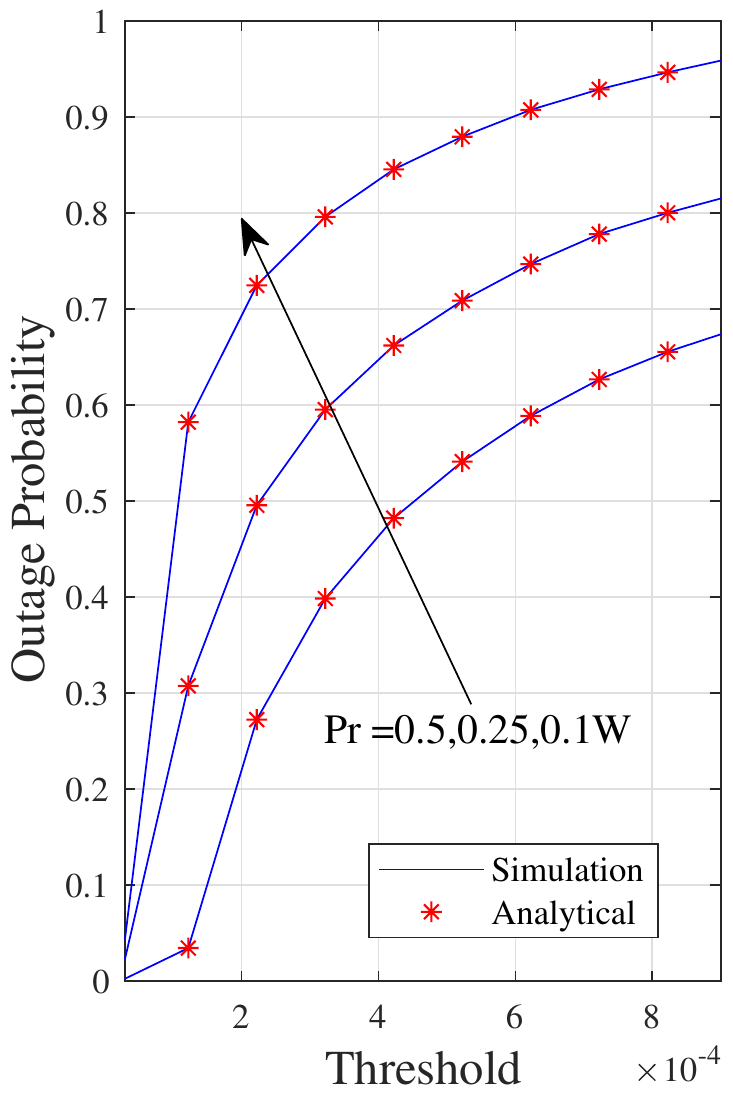}
\par\end{center}
\begin{center}
\caption{Outage probability versus the threshold value for different values
of relay transmit power.\label{fig:outagePr}}
\par\end{center}%
\end{minipage}
\end{figure}

The outage probability of the proposed system is plotted in Figs.
\ref{fig:OutageL} and \ref{fig:outagePr} versus the threshold for
different values of the height of the LED and the relay transmit power,
respectively. Looking at both figures, we can see that the performance
is considerably affected by these two factors as it degrades when
the relay transmit power increases and the vertical distance decreases.
On the other hand, it increases when the relay transmit power declines
and the height of the LED increases.

\section{\label{sec:Conclusion-1}Conclusions }

This paper analyzed the performance of the hybrid PLC/VLC system in
terms of the average capacity and outage probability for which analytical
expressions were derived and then verified by Monte Carlo simulations.
We investigated the effect of different system parameters on the performance
of the proposed system. The results revealed that the performance
of the proposed hybrid system is affected by various parameters, such
as the relay transmit power, the height of the LED and the maximum
cell radius of the LOS. It is worth pointing out that the use of such
hybrid systems can provide better mobility to the end user than that
provided by the PLC-only system.

\bibliographystyle{ieeetr}
\bibliography{references2}

\newcommand{\noopsort}[1]{} \newcommand{\printfirst}[2]{#1}
  \newcommand{\singleletter}[1]{#1} \newcommand{\switchargs}[2]{#2#1}
\begin{thebibliography}{10}

\bibitem{Leo2017}
L.~de~M.~B. A.~Dib, V.~Fernandes, M.~de~L.~Filomeno, and M.~V. Ribeiro,
  ``Hybrid {PLC}/wireless communication for smart grids and internet of things
  applications,'' {\em IEEE Internet Things. J}, vol.~PP, no.~99, pp.~1--1,
  2017.

\bibitem{T.Komine2003}
T.~Komine and M.~Nakagawa, ``Integrated system of white {LED} visible-light
  communication and power-line communication,'' {\em IEEE Trans. Consum.
  Electron.}, vol.~49, pp.~71--79, Feb. 2003.

\bibitem{Tan2011}
B.~Tan and J.~Thompson, ``Relay transmission protocols for in-door powerline
  communications networks,'' in {\em IEEE Int. Conf. Commun.(ICC)}, pp.~1--5,
  June. 2011.

\bibitem{D.lessandro2012}
S.~D'Alessandro and A.~M. Tonello, ``On rate improvements and power saving with
  opportunistic relaying in home power line networks,'' {\em EURASIP J.
  Advances. Signal Process.}, vol.~2012, p.~194, Sept. 2012.

\bibitem{Gupta17}
A.~Gupta, N.~Sharma, P.~Garg, and M.~S. Alouini, ``Cascaded {FSO-VLC}
  communication system,'' {\em IEEE Wireless Commun. Lett.}, vol.~6,
  pp.~810--813, Dec. 2017.

\bibitem{K.M.Rabie}
K.~M. Rabie and B.~Adebisi, ``Enhanced amplify-and-forward relaying in
  non-gaussian plc networks,'' {\em IEEE Access}, vol.~5, pp.~4087--4094, 2017.

\bibitem{K.R}
K.~M. Rabie, B.~Adebisi, H.~Gacanin, G.~Nauryzbayev, and A.~Ikpehai,
  ``Performance evaluation of multi-hop relaying over non-{G}aussian {PLC}
  channels,'' {\em J. Commun. Netw.}, vol.~19, pp.~531--538, Oct. 2017.

\bibitem{Rabie18}
K.~M. Rabie, B.~Adebisi, H.~Gacanin, and S.~Yarkan, ``Energy-per-bit
  performance analysis of relay-assisted power line communication systems,''
  {\em IEEE Trans. Green Commun. Networking}, vol.~2, pp.~360--368, Jun. 2018.

\bibitem{Zou2009}
H.~Zou, A.~Chowdhery, S.~Jagannathan, J.~M. Cioffi, and J.~L. Masson,
  ``Multi-user joint subchannel and power resource-allocation for powerline
  relay networks,'' in {\em IEEE Int. Conf. Commun.(ICC)}, pp.~1--5, Jun. 2009.

\bibitem{Moslem2016}
M.~Noori and L.~Lampe, ``Multi-way relaying for cooperative indoor power line
  communications,'' {\em IET Commun.}, vol.~10, no.~1, pp.~72--80, 2016.

\bibitem{Refik15}
O.~Narmanlioglu, R.~C. Kizilirmak, and M.~Uysal, ``Relay-assisted {OFDM}-based
  visible light communications over multipath channels,'' in {\em 2015 17th
  Int. Conf. Transparent Optical Netwr. (ICTON)}, pp.~1--4, Jul. 2015.

\bibitem{Tonello2012}
A.~M. Tonello, F.~Versolatto, B.~Bejar, and S.~Zazo, ``A fitting algorithm for
  random modeling the {PLC} channel,'' {\em IEEE Trans. Power Del.}, vol.~27,
  pp.~1477--1484, Jul. 2012.

\bibitem{Zego2009}
L.~Zeng, D.~C. O'Brien, H.~L. Minh, G.~E. Faulkner, K.~Lee, D.~Jung, Y.~Oh, and
  E.~T. Won, ``High data rate multiple input multiple output {MIMO} optical
  wireless communications using white led lighting,'' {\em IEEE Journal Sel.
  Areas Commun.}, vol.~27, pp.~1654--1662, December 2009.

\bibitem{GhethCSNDSPNew}
W.~Gheth, K.~M. Rabie, B.~Adebisi, M.~Ijaz, G.~Harris, and A.~Alfitouri,
  ``Hybrid power-line/wireless communication systems for indoor applications,''
  in {\em 2018 11th Int. Symp. Commun. Syst., Netw. and Digit. Signal Process.
  (CSNDSP)}, pp.~1--6, July 2018.

\bibitem{GHETH_GLOBECOM}
W.~Gheth, K.~M. Rabie, B.~Adebisi, M.~Ijaz, and G.~Harris, ``Performance
  analysis of integrated power-line/visible-light communication systems with af
  relaying,'' in {\em IEEE Global Commun. Conf. (GLOBECOM)}, 2018.

\end{thebibliography}
 
\end{document}